\documentstyle[prl,aps,floats,epsf]{revtex}
\newcommand{\beq}{\begin{equation}}
\newcommand{\eeq}{\end{equation}}
\newcommand{\beqa}{\begin{eqnarray}}
\newcommand{\eeqa}{\end{eqnarray}}
\newfont{\slb}{cmbxsl10 scaled \magstep5}

\newcommand{\Eq}[1]{(\ref{#1})}
\newcommand{\Fig}[1]{Fig.\ref{#1}}

\newcommand{\deffig}[4]{
\begin{figure}[tb]
  \begin{center}
    \hspace*{0em}
    \epsfxsize=#3\textwidth
    \null\ 
    \epsfbox{#2}
    \caption{#4}
    \label{#1}
  \end{center}
\end{figure}
}

\begin{document}

\draft

\twocolumn[\hsize\textwidth\columnwidth\hsize\csname @twocolumnfalse\endcsname

\title{Fractal Droplets in Two Dimensional Spin Glasses}
\author{Naoki Kawashima}
\address{
Department of Physics, Tokyo Metropolitan University,
Minami-Ohsawa 1-1, Hachiohji, Tokyo 192-0397, Japan
}
\date{\today}
\maketitle

\widetext

\begin{abstract}
The two-dimensional Edwards-Anderson model with Gaussian bond 
distribution is investigated at $T=0$ with a numerical method.
Droplet excitations are directly observed.
It turns out that the averaged volume of droplets is proportional to
$l^D$ with $D = 1.80(2)$ where $l$ is the spanning length of droplets,
revealing their fractal nature.
The exponent characterizing the $l$ dependence of the
droplet excitation energy is estimated as $\theta_D' = -0.42(4)$,
clearly different from the stiffness exponent for domain wall excitations.
\end{abstract}

\vskip1pc

]
\narrowtext

``{\it Droplet}'' is a well-known and useful concept in the study of
critical phenomena\cite{MFisher}. 
The droplet argument for spin glasses\cite{FisherHuse,FisherHuseFilms}
is based on this concept and is one of important working 
hypotheses in the field.
Its validity is being actively discussed, 
in particular, for the low-temperature phase
in three dimensions\cite{Controversy}.
Droplets are defined as collective excitations 
from some pure state below the transition temperature.
In the case of a typical spinglass model in two dimension, however,
the system is believed to be critical right at the zero-temperature.
Therefore, it is not clear, {\it a priori}, if we can treat
droplets defined at zero-temperature in two dimensions
in the same fashion as we treat the droplets below
the critical point in three dimensions.
In two dimensions,
most of the critical indices can be derived from a single exponent $\theta_D$
that characterizes the size-dependence of the droplet excitation energy,
according to the droplet argument or other similar arguments
such as the domain wall renormalization group theory\cite{BrayMoore84}.
For example, $-\theta_D$ was expected to be equal to $y_t \equiv 1/\nu$
while $-\theta_S$ was identified with $y_t$ \cite{BrayMoore84},
where $\theta_S$ is the stiffness exponent
which characterizes the system size dependence of the
domain wall excitation energy.
A deviation from this standard story for two-dimensional spin glass models
was first reported based on numerical calculation of magnetization
at zero-temperature with finite field\cite{KawashimaI}, 
in which we found that the thermal exponent $y_t$ does not agree with 
$-\theta_S$.
Several subsequent numerical works
\cite{KawashimaII,Liang92,Rieger97,NifleYoung97}
confirmed this result.
Thus the equivalence among three exponents,
$y_t, -\theta_S$ and $-\theta_D$, 
should now be a subject to a close re-examination.
As for the equivalence between $y_t$ and $-\theta_D$,
so far no evidence of its violation has been found.
However, this may be simply because identifying droplets is
technically so difficult that nobody has actually succeeded in
direct observation of droplets large enough to discuss asymptotic behavior.
In fact, if one adopts the original definition\cite{FisherHuse} of 
droplets and apply a simple combinatorial optimization algorithm, 
it would be impossible to obtain droplets larger 
than a few lattice spacings.

In this letter, we focus our attention 
on the EA model on a square lattice
$$
  {\cal H} \equiv - \sum_{(ij)} J_{ij} S_i S_j
$$
where $J_{ij}$'s are quenched Gaussian random variables 
with the mean value of zero and the standard deviation of $J$.
We take $J$ as the unit of energy throughout the present letter.
It should be noted that the one-parameter scaling which was
derived from various ``pictures'' such as the droplet picture
and the domain wall renormalization group argument
can be also derived within the framework of standard 
finite size scaling theory without any assumptions
or pictures.
We start with the following standard form for the partition function,
\begin{equation}
  \log \hat Z(T,H,L) \sim f(TL^{y_t}, HL^{y_h})
  \label{eq:FSS}
\end{equation}
where $\hat Z, T, H, L$ are the singular part of the partition function,
temperature, magnetic field, and system size, respectively.
We hold \Eq{eq:FSS} to be a defining equation of
the exponent $y_t$.
One of the two scaling exponents can be fixed by using the fact that 
the magnetization at zero temperature and zero magnetic field
is proportional to $L^{d/2}$ because of the absence of non-trivial
degeneracy.
This leads to $y_h = y_t + d/2$.
In previous works, we measured the magnetization 
$M(T=0,H,L)$\cite{KawashimaI} and the spin-glass susceptibility
$\chi_{\rm SG}(T,H=0,L)$\cite{KawashimaII}
which are both derivatives of $\log \hat Z$ with respect 
to $T$ and $H$ multiplied by some known factors.
By matching the numerical results with the asymptotic form
derived from \Eq{eq:FSS}, we obtained $y_t \simeq 0.48$
in both calculations.

On the other hand, it is often assumed, or for some models proved, 
that the excitation energy $E_{W}$ of domain walls induced by twisting
the system has the same scaling behavior as the singular part 
of the free energy.
This means $\theta_S = -y_t$ where $\theta_S$ is defined by
$$
  E_{W}(L) \propto L^{\theta_S} \quad (T=0).
$$
In the early stage of the study on the present issue,
the domain wall energy was measured for various sizes\cite{BrayMoore84} 
with the hope of obtaining an estimate of $y_t$.
The stiffness exponent was estimated as $-\theta_S \simeq 0.29$
\cite{BrayMoore84}.
This result was reconfirmed later by other
groups\cite{Rieger97,KawashimaAoki99} with larger scale computations.
Therefore, it is now rather clear that $-\theta_S$ and $y_t$
are two different exponents.

Fisher and Huse defined\cite{FisherHuse}
a droplet of the scale $\lambda$ including a given site $i$ as
a cluster of spins (with $i$ among them) with the smallest 
excitation energy that contains more than $\lambda^d$ and 
less than $(2\lambda)^d$ spins.
The basis of the argument then is the following scaling 
form that describes the excitation-energy distribution of droplets of
the scale $\lambda$:
\begin{equation}
  P_{\lambda}(E_{\lambda}) = 
  \frac{1}{\Upsilon {\lambda}^{\theta_D}}
  \tilde P\left(\frac{E_{\lambda}}{\Upsilon {\lambda}^{\theta_D}}\right)
  \label{eq:DropletEnergyScaling}
\end{equation}
where $\Upsilon$ is some constant, $\theta_D$ is the droplet exponent, 
and $\tilde P(x)$ is the scaling function which is continuous and 
non-vanishing at $x=0$.
Since the droplet ``size'', $\lambda$, is defined to be 
proportional to (volume)$^{1/d}$,
it is not necessarily a spanning length of droplets 
because the volume of a droplet may in general 
have a non-trivial fractal dimension.
In fact, in what follows we demonstrate that this is the case, i.e.,
a typical large droplet in two dimensions occupies only an 
infinitesimal fraction of the volume of the box that contains it.
Therefore, for a given droplet, we have at least two different length 
scales, $\lambda$ and the spanning length.
We refer hereafter to the fractal dimension 
of the droplet volume as $D$.
Then, we have the relation 
$$
  V = \lambda^d \propto l^D
$$
where $l$ is the spanning length of the droplet.

Since the above-mentioned definition of droplets
is inconvenient from computational point of view,
we have adopted the following alternative definition.
First we consider an $L\times L$ system with free boundary condition.
We define a droplet of scale $L$ as the cluster of spins
that has the smallest excitation energy among those which
contain the central spin and contains no spins on the boundary.

If we assume the conventional droplet argument,
the central spin is surrounded by droplets (in the original definition) 
of various scales.
Since perimeters of smaller droplets are far from the boundary of the system
they must be relatively free from the influence of the constraint imposed
on the boundary.
Therefore, smaller droplets in a finite system should have the same
scaling properties as they would in an infinite system.
On the other hand, the fixed boundary imposes
strong restrictions upon the shapes that larger droplets can take.
As a result, they have larger excitation energy 
than they would in an infinite non-restricted system.
Therefore, a crossover size of droplets must exist and it separates
the region of smaller sizes in which one observes a correct asymptotic behavior
from the other region of larger sizes that is strongly affected by the boundary.

\deffig{fg:Examples}{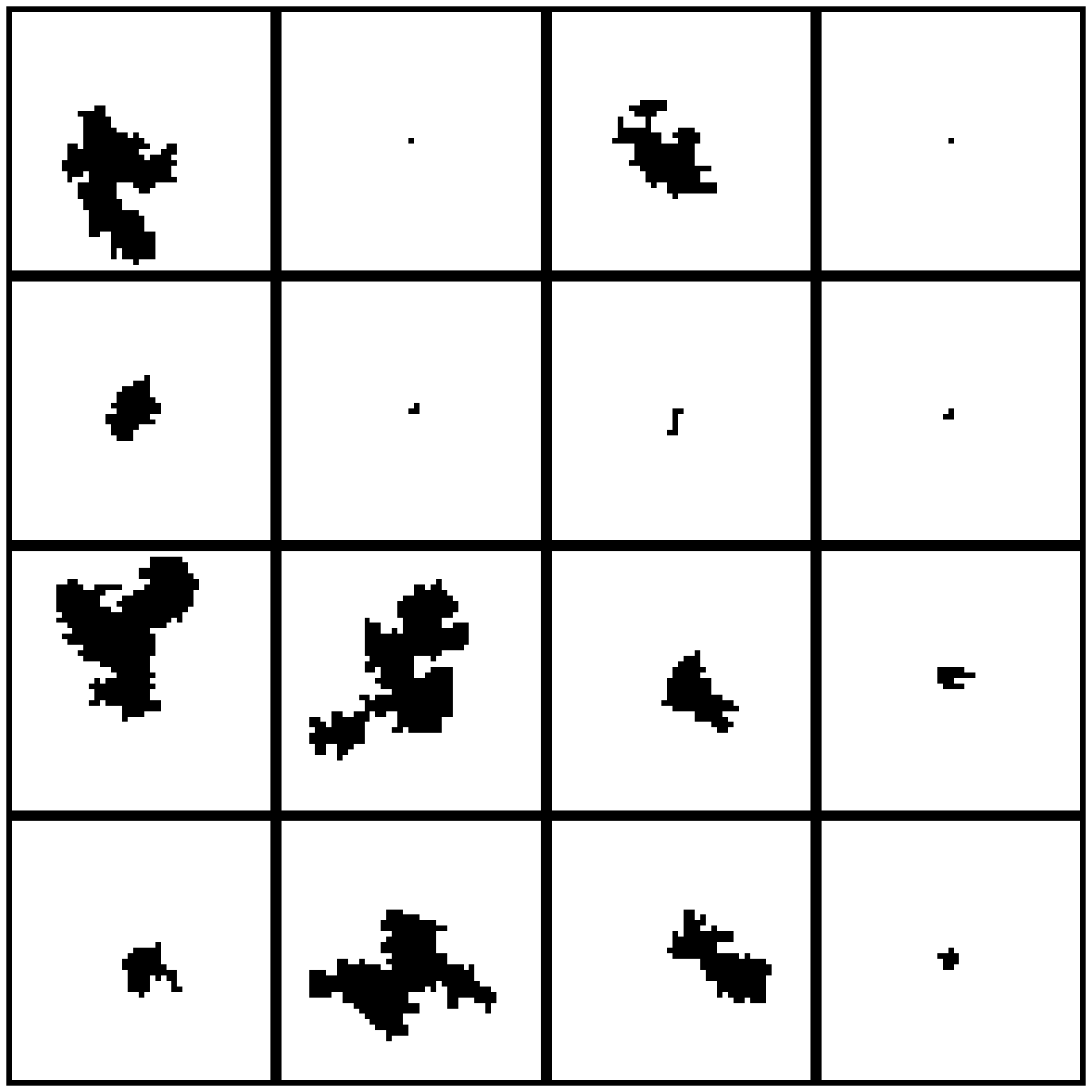}{0.38}{16 randomly chosen droplets.}
In the spirit of effective mutual independence among the droplets
of different scales discussed in \cite{FisherHuse},
it is natural to expect that the crossover size corresponds to
the largest droplet that can fit into a box of $L\times L$, i.e.,
the droplet whose spanning length is of the order of $L$.
Then, considering the fact that larger droplets tend to have
smaller excitation energy,
one may also expect that the spanning length of the droplet identified 
by the procedure described above would be proportional to $L$.
We here define $\theta_D'$ by the following $L$ dependence of 
the average excitation energy of the droplets in our definition,
$$
  [E_D(L)] \propto L^{\theta_D'},
$$
where $[\cdots]$ stands for the average over samples.
Comparing this to the original definitions of the droplet size 
and droplet exponent, 
we obtain
\begin{equation}
  \theta_D' = \frac{D}{d}\theta_D. \label{eq:Relationship}
\end{equation}
because $L$ is proportional to the spanning length of the droplet.

In order to observe droplets in the new definition,
we first compute the ground state with free boundary condition,
and take it as the reference spin configuration.
Then we compute the ground state with the constraint that
the spins on the boundary are to be fixed as they are in the reference state 
while the central spin is to be fixed opposite, 
thereby forcing a cluster of spins including the central spin to flip.
For a system with the free boundary condition, 
polynomial-time optimization algorithms are available 
whereas for the systems with constraints no such algorithm is known.
In fact, the two dimensional spin glass problem with general 
constraints has been proven to be NP hard\cite{Barahona82}.
Therefore, we have employed the replica optimization\cite{KawashimaI},
which is a heuristic optimization algorithm based on the idea of
renormalization group.
The details of the algorithm are described elsewhere
\cite{KawashimaI,KawashimaFuture}.

Since the algorithm is only heuristic,
we need to estimate the rate of obtaining false ground states and
how much such states may affect the average values of
various quantities of interest.
For this purpose, we compared our results with exact solutions 
when they are available.
We have actually found no error with the settings of algorithmic 
parameters that are used in the computation presented in this letter.
We detected some errors only when we loosened the termination condition.
Even in that case, the rate and the amount of the errors turned out 
to be so small that their effect is well within the final statistical errors.
Therefore, we believe that the errors are negligible even for 
the largest systems ($L=49$), for which no exact result is 
available for comparison.


%
%
In \Fig{fg:Examples}, we present 16 randomly chosen droplets
in the system of size $L=49$.
We can see that the size of the droplets varies depending on the sample.
Some of them contains only one spin.
This reflects the fact that $|\theta_D|$ is small
and the excitation energy does not very strongly depend on the droplet size.
Frequency of having small droplets decreases as a function of the system size.
On the other hand, we also see that larger droplets whose size is 
comparable even to the system itself occupy a considerable fraction 
in the whole population of droplets.
In these larger droplets, many fingers and overhangs
can be observed, which already suggests the fractal nature of the droplets.

In order to check that the spanning length of a typical droplet
is really proportional to the system size as we expect,
we compute two length scales $l_0$ and $l_2$.
The length $l_0$ is the spanning length itself, i.e.,
the difference in $x$-coordinates of the left-most site 
and that of the right-most site in a droplet,
assuming that the $x$-axis stretches from left to right.
On the other hand, the length $l_2$ is defined as
$$
  l_2 \equiv \sqrt{\langle(x-\langle x \rangle)^2\rangle}.
$$
Here,
$$
  \langle x^n \rangle \equiv
  \left. \sum_{i \in \mbox{droplet}} x_i^n \right/
  \sum_{i\in\mbox{droplet}} 1
$$
where $x_i$ is the $x$-coordinate of the site $i$.
\deffig{fg:LengthScales}{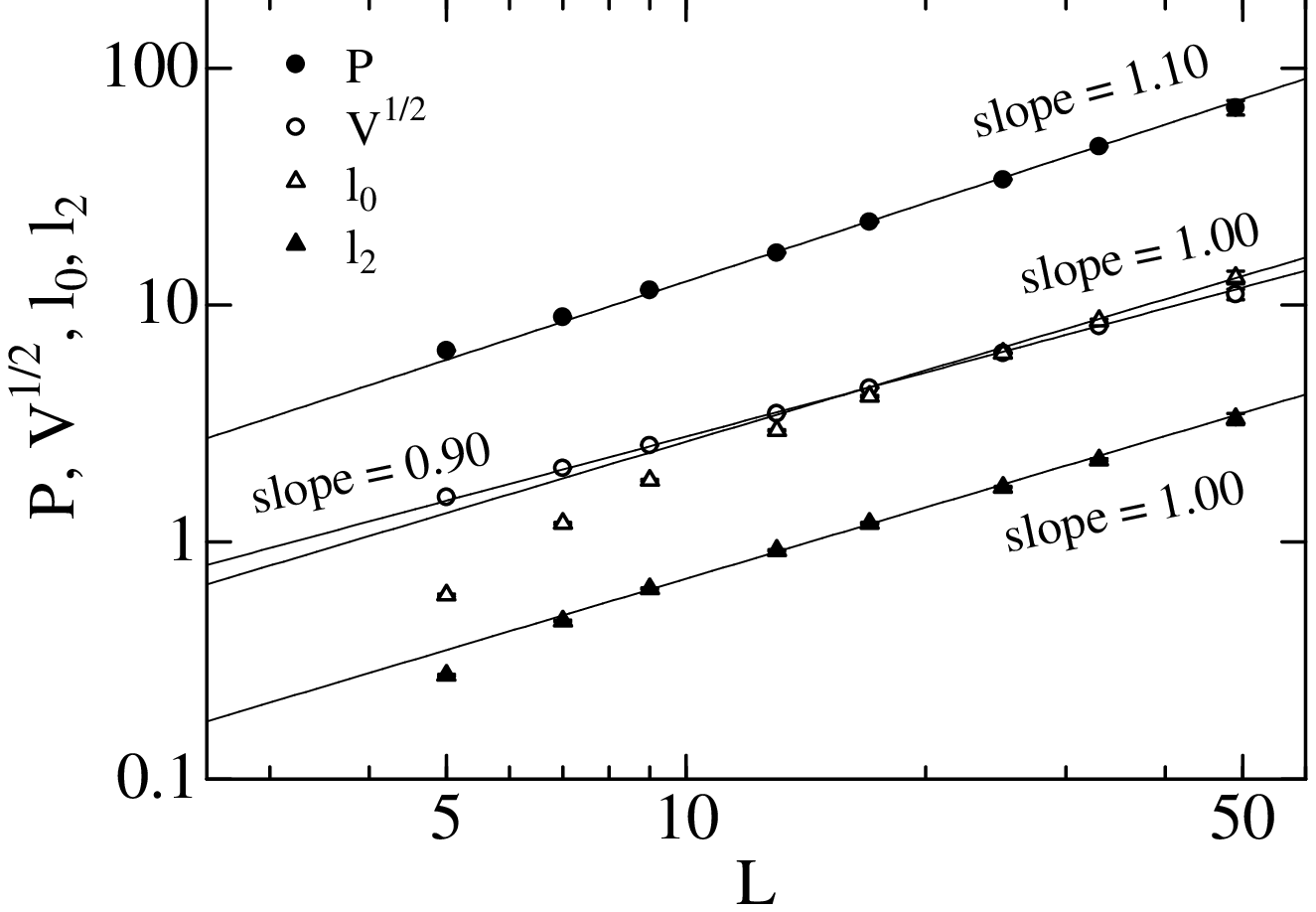}{0.42}{
Various length scales of droplets.
The statistical errors for $L=49$ are comparable to the symbol size
whereas those for other data points are much smaller than the symbols.
}
In \Fig{fg:LengthScales}, these two length scales averaged 
over many samples are shown as a function of the system size $L$.
A correction to scaling can be observed for small system sizes $L \le 17$.
In particular the slope for $l_0$ may seem
slightly larger than unity at a first glance.
However, since the slope cannot asymptotically exceed unity 
for an obvious reason,
we consider that $l_0$ is proportional to $L$ in the large $L$ limit.
The linearity for $l_2$ is much better and indicates proportionality 
to $L$.

The averaged length, $P$, of the boundary of the droplet and
the averaged volume, $V$ are also plotted in \Fig{fg:LengthScales}.
The linearlity in the logarithmic scale is rather good for both
$P$ and $V^{1/2}$ and both of them have slopes different from unity.
For $P$, we obtain $ P \propto L^{D_s} $
with the surface fractal dimension
$$
  D_s = 1.10(2).
$$
For $V$, we estimate the fractal dimension as
\begin{equation}
  D = 1.80(2). \label{eq:D}
\end{equation}
The fractal dimension $D$ of droplets is certainly smaller than $d=2$.
Thus, we conclude that the droplets at the critical point $T=0$
have fractal nature in the volume as well as in the perimeter.

%
Finally, we measure the droplet excitation energy $E_D(L)$.
Similar to the size and the shape, the droplet excitation energy
has a broad distribution.
When rescaled with the average value $E_D(L)$,
the histograms of excitation energies
for various system sizes fit on top of each other, 
showing the validity of the form \Eq{eq:DropletEnergyScaling}
(See \Fig{fg:DropletEnergyScaling}).
In addition, we observe that the scaling function $\tilde P(X)$ 
has a non-vanishing value at $X=0$, satisfying a necessary condition
for the droplet argument to be valid.

%
Since the average droplet excitation energy does not strongly depend on
the system size, it is subject to relatively large correction
to scaling and the estimate of $\theta_D'$ can be affected by the deviation
from the asymptotics in small $L$ region.
Therefore, the estimate depends on which we choose among 
the above-mentioned length scales that differ from each
other for small sizes.
When $\log E_D$ is plotted against $\log L$ the slope is estimated
to be $\theta_D' \sim -0.42$, whereas it becomes $-0.38$ and $-0.45$
when $l_0$ and $l_2$, respectively, are chosen instead of $L$.
These three plottings are shown in \Fig{fg:DropletEnergy} together
with straight lines with slopes of corresponding estimates of $\theta_D'$.
From these results we quote the following value as the estimate
of the droplet exponent $\theta_D'$:
\begin{equation}
  -\theta_D' = 0.42 (4). \label{eq:DropletExponent}
\end{equation}
Assuming \Eq{eq:Relationship} and using the value of $D$ in \Eq{eq:D},
we obtain the following estimate of $\theta_D$,
\begin{equation}
  -\theta_D = 0.47(5)
\end{equation}
in good agreement with previous estimates of $y_t$ such as
$y_t = 0.48(1)$~\cite{KawashimaII}.
\deffig{fg:DropletEnergyScaling}{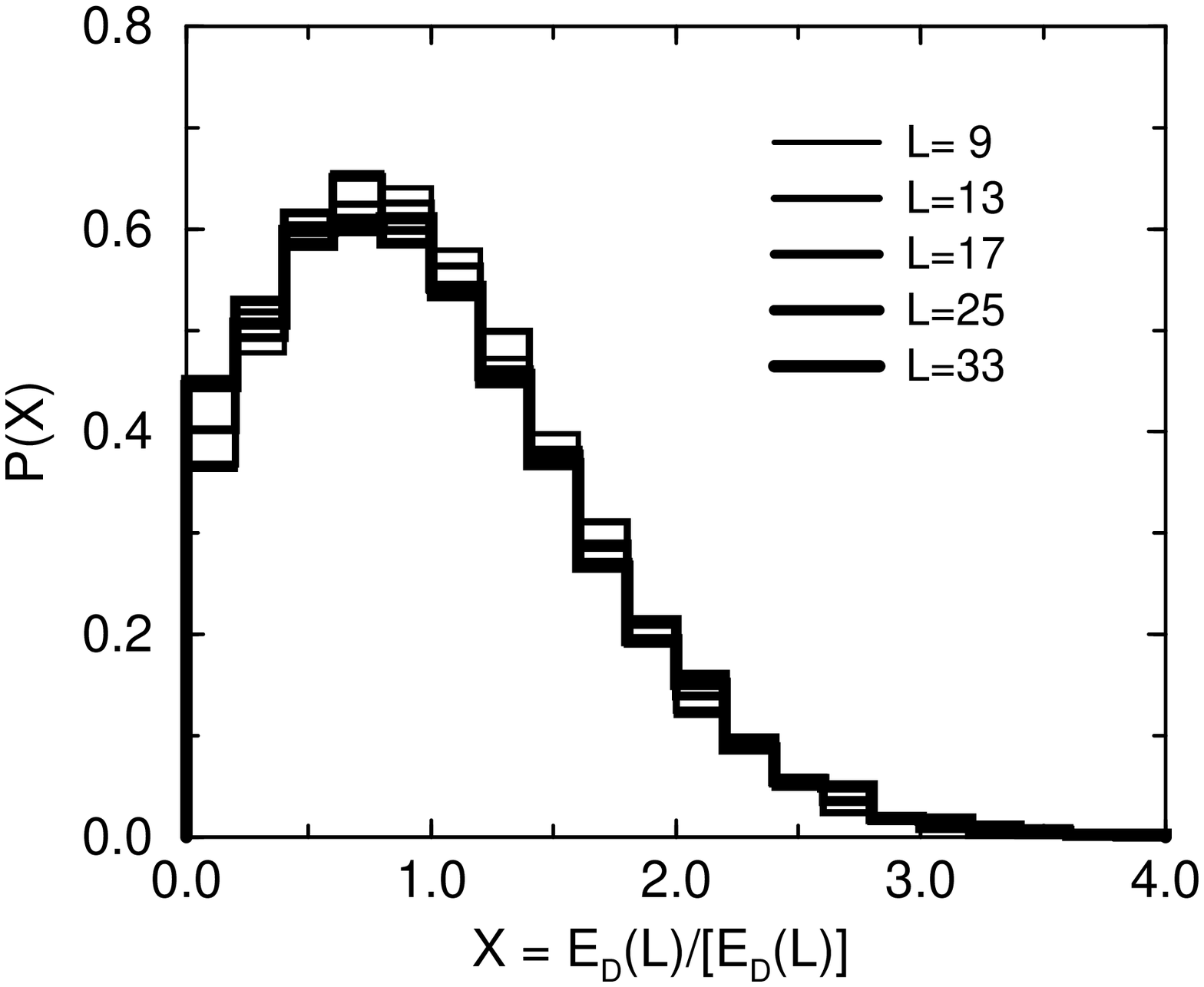}{0.46}
{The distribution of droplet excitation energies 
rescaled with the average values.}
\deffig{fg:DropletEnergy}{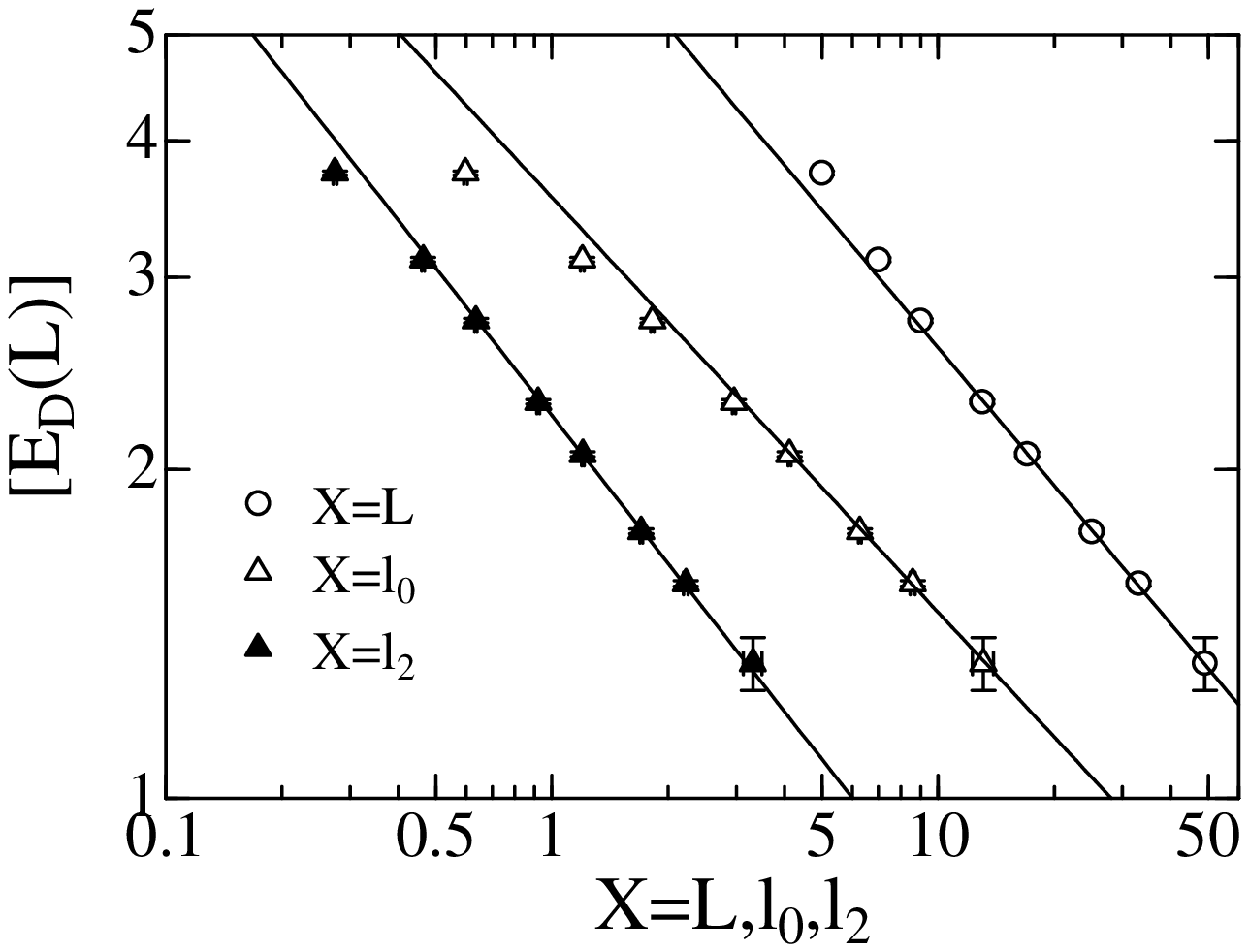}{0.42}
{The average droplet excitation energy.}

To summarize, we have generated droplets in two dimensions
for the EA model with Gaussian bond distribution.
We have found the droplets non-compact in contrast 
to what is assumed in the conventional droplet argument for
higher dimensions.
The exponent $-\theta_D'$ has been estimated to be slightly smaller
than the thermal exponent in the absolute value, and this difference
can be well understood within the framework of the finite-size scaling
and the standard droplet argument, by taking into account the fractal 
nature of droplets.
It is rather surprising that the droplet argument seems to be valid 
for two dimensional critical behavior in spite of the qualitative 
difference between fractal droplets at the critical point shown
in the present letter and the off-critical, presumably compact,
droplets in higher dimensions. 
The disagreement between the droplet exponent and
the stiffness exponent also remains puzzling.
At least, however, the present result suggests that
a domain wall of a $L\times L$ system may not necessarily be considered
as an object of scale $L$ because of the existence of different 
ways for measuring the size of the same object.
In other words, at present we do not know which way of measuring
we should use for comparing the size of a domain wall
and that of a droplet.

The author thanks 
A.~P.~Young and H.~Rieger for valuable comments.
He is also grateful to 
T.~Aoki for his technical assistance.
This work is supported by Grant-in-Aid for Scientific Research Program
(No.11740232) from the Ministry of Education, Science, Sport and Culture
of Japan.

\end{document}